\documentclass[aps,prd,superscriptaddress,showpacs,nofootinbib,floatfix,preprint]{revtex4}

\usepackage{graphicx}	% Include figure files

\newcommand{\beq}{\begin{equation}}
\newcommand{\eeq}{\end{equation}}
\newcommand{\beqa}{\begin{eqnarray}}
\newcommand{\eeqa}{\end{eqnarray}}
\newcommand{\bpr}{\begin{problem}}
\newcommand{\epr}{\end{problem}}
\newcommand{\bcent}{\begin{center}}
\newcommand{\ecent}{\end{center}}
\newcommand{\bfig}{\begin{figure}}
\newcommand{\efig}{\end{figure}}
\newcommand{\bpc}{\begin{picture}}
\newcommand{\epc}{\end{picture}}
\newcommand{\barr}{\begin{array}}
\newcommand{\earr}{\end{array}}
\newcommand{\bitm}{\begin{itemize}}
\newcommand{\eitm}{\end{itemize}}
\newcommand{\bright}{\begin{flushright}}
\newcommand{\eright}{\end{flushright}}
\newcommand{\bminip}{\begin{minipage}}
\newcommand{\eminip}{\end{minipage}}
\newcommand{\btab}{\begin{tabular}}
\newcommand{\etab}{\end{tabular}}

\newcommand{\hiroshima}{Graduate School of Science, Hiroshima University, Kagamiyama, Higashi-Hiroshima 739-8526, Japan}
\newcommand{\izest}{International Center for Zetta-Exawatt Science and Technology, Ecole Polytechnique, Route de Saclay, Palaiseau, F-91128, France}
\newcommand{\gist}{Center for Relativistic Laser Science, Institute for Basic Science (IBS), Gwangju 500-712, Republic of Korea}

\begin{document}

%Title of paper
\title{
Testing helicity dependent $\gamma\gamma\rightarrow\gamma\gamma$ 
scattering in the region of MeV
}

\author{K. Homma} \affiliation{\hiroshima}\affiliation{\izest}
\author{K. Matsuura} \affiliation{\hiroshima}
\author{K. Nakajima} \affiliation{\gist}

\date{\today}

\begin{abstract}
Light-by-light scatterings contain rich information on the photon coupling 
to virtual and real particle states.
In the context of quantum electrodynamics (QED),
photons can couple to a virtual $e^+e^-$ pair.
Photons may also couple to known resonance states in the context of
quantum chromodyanmics and electroweak dynamics 
in higher energy domains and possibly couple to
unknown resonance states beyond the starndard model.
The perturbative QED calculations manifestly 
predict the maximized cross section at the MeV scale, however, any example 
of the exact real-photon - real-photon scattering has not been observed hitherto. 
Hence, we propose the direct measurement with the maximized cross-section 
at the center-of-mass system energy of 1-2 MeV to establish the firm footing 
at the MeV scale. Given currently state-of-the-art high power lasers, 
the helicity dependent elastic scattering may be observed at a reasonable rate, 
if a photon-photon collider exploiting $\gamma$-rays generated by 
the inverse nonlinear Compton process with electrons delivered from laser-plasma 
accelerators (LPA) are properly designed. 
We show that such verification is feasible in a table-top scale
collider which may be an unprecedented breakthrough in particle accelerators 
for basic physics research in contrast to energy frontier colliders.
\end{abstract}
% 12.20.-m	Quantum electrodynamics
% 12.20.Fv	Experimental tests (for optical tests in quantum electrodynamics, see 42.50.Xa)
% 14.80.Va	Axions and other Nambu-Goldstone bosons (Majorons, familons, etc.)
%29.20.db　Storage rings and colliders
%41.75.Jv　Laser-driven acceleration
%42.62.-b　Laser applications.

\maketitle

\section{Introduction}
Light-by-light scattering is a purely quantum process, hence, in the standard model, 
only possible in the context of quantum electrodynamics (QED). 
QED is the most strictly tested dynamics in its perturbative regime. 
The scattering amplitude in the lowest order is described by the box diagram 
where a virtual $e^+e^-$ pair loops between four external photons.
So far the diagram is indirectly confirmed with off-shell external photons via 
$\gamma A\rightarrow \gamma A$~\cite{DelbrueckExp} 
known as Delbr\"uck scattering~\cite{Delbrueck} and 
also proposed to be further tested via $A A \rightarrow A A\gamma \gamma$ 
in the higher energy domain at the Large Hadron Collider~\cite{LHCLbyL}. 
In these scatterings, however, photons are emitted from nuclei $A$ and 
have finite off-shell masses. With the exact on-shell condition, namely, 
real-photon - real-photon scattering has not been demonstrated hitherto 
despite of its explicit predictions~\cite{Halpern:1934,Toll:1952,R5,R6,R7}. 
Moreover, its dependence on the photon polarization states
has not been tested at all. By means of off-shell incident photons from nuclei, 
it is difficult to test the polarization property. 
Photon-photon interactions also contain rich information on 
the two-photon coupling to standard model / non-standard model 
resonance states depending on the center of mass energy. 
In the MeV range, no direct search for resonance states
coupling to two photons has been performed.
Therefore, it is indispensable to first verify the purely QED-based 
scattering amplitude 
with specified polarization states in a pristine experimental condition,
because a significant deviation from the prediction indicates 
the existence of physics beyond the standard model. 

\begin{figure}
\includegraphics[width=0.7\linewidth]{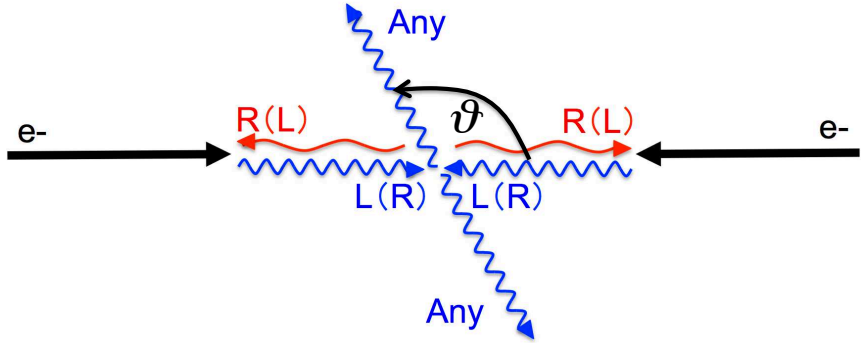}
\caption{ 
Helicity configuration in $\gamma\gamma$ scattering via inverse Compton
scatterings between cicularly polarized laser pulses and 
unpolarized electrons. 
R and L denote right- and left-handed helicity
states of photons, respectively.
}
\label{Fig0}
\end{figure}

In lower energies, searches for photon-photon
interactions have been performed at keV~\cite{R8}, eV~\cite{R9,R10} and 
sub-eV energy scales attempting to search for dark matter
~\cite{R11,R12,R13,R14,R15,R16,R17,R18,R19} and also 
to put the upper limit on the QED photon-photon interaction~\cite{R20}. 
There are several proposals~\cite{Lundstrom,Heinzl:2006xc,R21,R22} 
to probe photon-photon interactions 
in the context of the low energy limit of the QED interactions
~\cite{Euler:1935zz,Heisenberg:1935qt} based on only optical photons.
These proposals commonly suffer from its 
extremely weak interaction. Therefore, currently state-of-the-art high-power 
lasers are supposed to be indispensable. However, if such lasers are available,
we may also use them to generate $\gamma$-rays. This idea leads us to try to 
consider the direct measurement of the scattering with the maximized 
cross-section at the center-of-mass system energy of 1-2 MeV 
rather than based on 1 eV photons.

In this paper we discuss the feasibility to measure
the helicity dependent $\gamma\gamma$-scattering
based on the helicity configuration as illustrated in Fig.\ref{Fig0}
where circularly polarized laser pulses are reflected upon unpolarized
electron bunches and $\gamma$-rays in the same helicity state collide head-on. 
We then consider to measure the elastic scattering 
without specifying the helicity states of the final state $\gamma$-rays. 
In contrast to the helicity dependence emphasizing on the forward scattering
amplitude~\cite{Dinu:2013gaa,Dinu:2014tsa}, we rather aim
at larger angle scattering events in order to verify 
the purely perturbative prediction. In the following paragraphs, we discuss 
the QED-based helicity dependent $\gamma\gamma$-scattering cross section, 
how to produce electrons with LPA, the expected helicity specified 
$\gamma$-ray yield based on the nonlinear Compton process and then 
possibility to realize a realistic experiment. 

\begin{figure}
\includegraphics[width=0.7\linewidth]{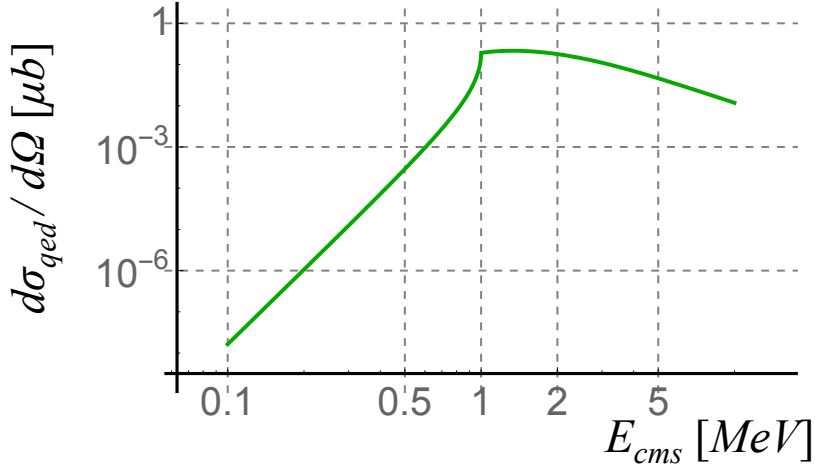}
\caption{ The unpolarized QED-based differential cross section 
$d\sigma_{qed}/d\Omega$ at $\vartheta = \pi/2$ 
in units of $\mu$b as a function of 
the center-of-mass system energy in photon-photon collisions,
$E_{cms}$.
}
\label{Fig1}
\end{figure}

\begin{figure}
\includegraphics[width=0.6\linewidth]{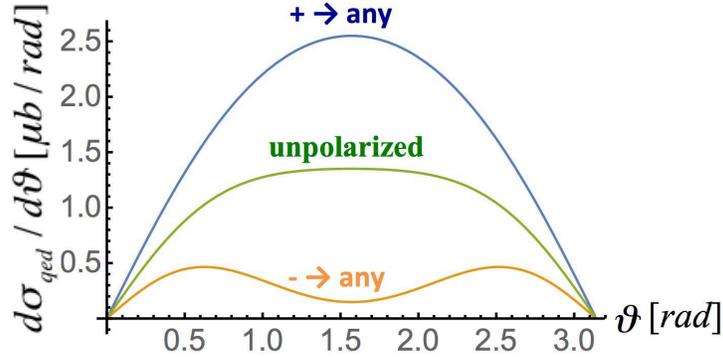}
\caption{ 
Helicity dependence of QED-based differential cross sections 
at $E_{cms}=1.4$ MeV as a function of the photon scattering angle $\vartheta$
in CMS. 
}
\label{Fig1b}
\end{figure}

\section{QED-based photon-photon interaction}
The elastic scattering of photons by photons has been thoroughly evaluated 
in \cite{R5,R6,R7}.
The differential cross section per solid angle 
for the unpolarized photon-photon scattering 
in the lower energy limit
% in $E_{cms}/m_e \ll 1$
is approximated as
%\beqa\label{eq_1}
%\frac{d\sigma_{qed}}{d\Omega} \sim \frac{\alpha^2 {r_0}^2}{4\pi^2}
%\frac{139}{(90)^2}\left(\frac{k}{m}\right)^6 (3+\cos\vartheta)^2 \\ \nnb
%\times
%\left(1+\frac{160}{139}\frac{(k/m)^2\sin^2\vartheta}{3+\cos^2\vartheta}\right),
%\eeqa
$
\frac{d\sigma_{qed}}{d\Omega} \sim \frac{\alpha^2 {r_0}^2}{4\pi^2}
\frac{139}{(90)^2}\left(\frac{k}{m}\right)^6 (3+\cos\vartheta)^2,
$
where
$m$ is electron mass in units of $\hbar=c=1$,
$k$ is the incident photon energy in the center-of-mass system,
$\alpha = 1/137$ is the dimensionless fine structure constant,
$r_0 = \alpha/m \sim 2.8 \times 10^{-13}$~cm is
the classical electron radius, and
$\vartheta$ is the polar angle of scattered photons with respect to
the colliding axis between two incident photons in the center-of-mass 
system on the reaction plane.
For laser photons of $k \sim 1$~eV, the total cross section is 
$\sigma \sim 10^{-42}$~b.
This is extremely small due to the steep $k^6$ dependence.
This situation is shown in Fig.\ref{Fig1} where the differential
cross section of the unpolarized photon-photon elastic scattering,
$d\sigma_{qed}/d\Omega$ at $\vartheta = \pi/2$, is plotted as a function of 
the center-of-mass system energy of photon-photon collisions, $E_{cms}$.
The cross section curve is obtained from the numerical 
calculation applying formulae in Ref.~\cite{R7}.
On the other hand, if $k = 0.5-1.0$~MeV is realized, 
the total cross section is maximized up to $\sigma \sim 1$~$\mu$b.
Therefore, in order to detect the perturbative QED
cross section, it is reasonable to perform the scattering experiment 
tuned at that energy range.
In the region $E_{cms} = 1-2$~MeV, we see a flat-top character.
This allows relatively large fluctuations on $E_{cms}$, which is preferable
for the photon-photon collider exploiting $\gamma$-rays 
via laser Compton scattering off electrons from LPA having a percent-level energy spread.

Figure \ref{Fig1b} shows the helicity dependent QED-based differential cross section
$d\sigma^{\pm \rightarrow any}/d\theta$ compared to the unpolarized cross section 
where the symbols $\pm \rightarrow any$ indicate the same and opposite 
helicity states between incident two $\gamma$-rays 
going to all the possible helicity states. 
These predictions are results of numerical calculations at $E_{cms}=1.4$ MeV 
based on Ref.~\cite{R7}.
We find a larger cross section in the same helicity incidence case
particularly around $\vartheta \sim \pi/2$. 

\begin{figure}
\includegraphics[width=1.0\linewidth]{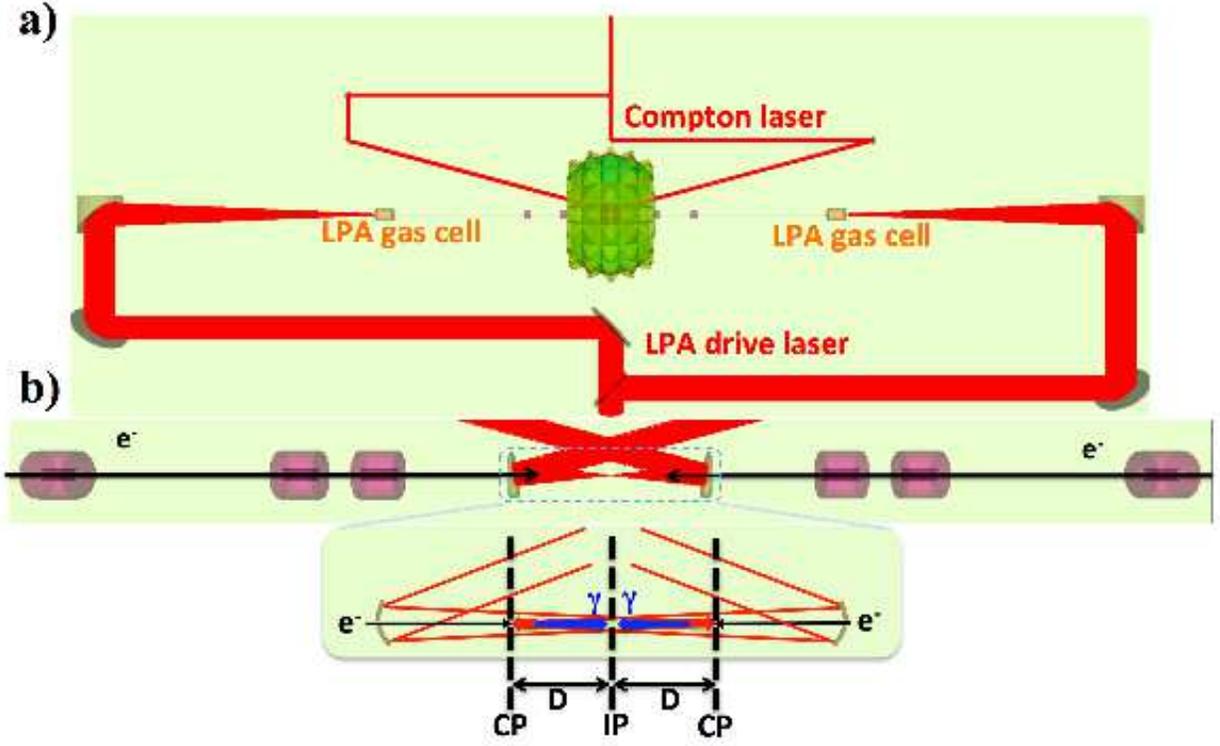}
\caption{ 
An all-optical table-top (3.4 m $\times$ 1.3 m) $\gamma\gamma$-collider: 
a) top-view including two LPAs and the detector system to capture 
the $\gamma\gamma\rightarrow\gamma\gamma$ scattering
and
b) collision geometry around the interaction point, IP, 
where $\gamma$-rays are produced at each Compton scattering point (CP) 
in head-on collisions and $D$ is the distance between IP and CP. 
}
\label{Fig2}
\end{figure}

\section{Table-top $\gamma\gamma$-collider}
Figure \ref{Fig2} a) shows a table-top (3.4 m $\times$ 1.3 m) $\gamma\gamma$-collider 
with a realistic detector system to capture large angle 
$\gamma\gamma\rightarrow\gamma\gamma$ scattering events. 
The system consists of two LPAs to generate electron beams with 
which two incident $\gamma$-beams are further produced via the inverse 
Compton process in head-on collisions. Initially synchronized two laser pulses 
are incident from the top and bottom sides of the top view, respectively. 
They are individually split into two drive pulses for LPA and scatter pulses 
for the successive inverse Compton process with electrons delivered from LPA. 
Figure \ref{Fig2} b) illustrates the collision geometry around the interaction point (IP)
of $\gamma\gamma$-scattering, which is located at the distance $D$ 
from the inverse Compton scattering point (CP).

\begin{figure}
\includegraphics[width=1.0\linewidth]{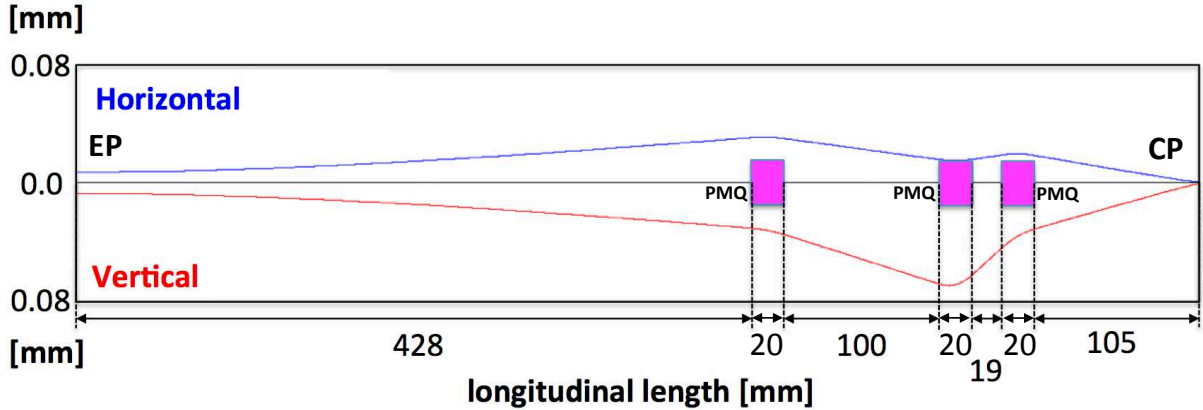}
\caption{ 
A focusing system for an electron beam produced by LPA consisting of 
three PMQs from the ejection point (EP) of the electron beam at 
the gas cell for LPA to the inverse Compton scattering point (CP). 
The blue and red curves are traces of beam envelopes simulated 
by TRACE3D~\cite{R30} for horizontal and vertical directions, respectively.
}
\label{Fig3}
\end{figure}

An electron beam with 210 MeV and 1.6 nC are produced from 
a two-stage laser wakefield accelerator~\cite{R23} comprising a 5-mm long gas 
cell filled with the mixed gas (e.g., 94 \% helium and 6 \% nitrogen) for 
the injector stage and a variable-length gas cell filled with pure helium 
for the accelerator stage. 
Designing parameters of the laser wakefield accelerator can be carried out by 
relying on the scaling law of nonlinear plasma wakefields 
in the bubble regime~\cite{R24,R25,R26}. 
Provided that a laser pulse with 41 TW peak power and 85 fs duration is 
focused on 12 $\mu$m spot radius on the entrance of the injector cell operated 
at plasma density of $3.3\times 10^{18}$ cm${}^{-3}$, 
strong nonlinear wakefields can be generated so that a 1.6 nC electron bunch 
could be trapped due to ionization-induced injection~\cite{R27,R28}
and accelerated up to 40 MeV, followed by boosting its energy up to 210 MeV 
at the length of 2.6 cm in the accelerator cell operated at plasma density 
of $1.1 \times 10^{18}$  cm${}^{-3}$. The relative energy spread and 
the normalized emittance of resultant output beams are estimated to 
be 4 \% in r.m.s. and 0.15 mm mrad, respectively.
  
The electron bunch is then focused via a set of permanent-magnet-based 
quadrupoles (PMQs)~\cite{R29} consisting of three elements 
-275 T/m, 770 T/m and -650 T/m with the common inner radius 3.0 mm and 
outer radius 12.0 mm, respectively, over 71.2 cm 
as shown in Fig.\ref{Fig3} 
which displays variations of the horizontal and vertical beam envelopes 
simulated by TRACE3D~\cite{R30} 
for the given incident parameters of the LPA configuration above. 

Based on these counter-propagating electron beams,
we evaluate the number of incident $\gamma$-rays produced by the inverse nonlinear
Compton process between a circularly polarized laser pulse and an unpolarized
electron bunch produced by LPA. 
It is worth noting that once we fix the focusing geometry, 
we cannot increase the effective number of $\gamma$-rays as much as we like
even if we could increase the laser pulse energy. This is because of
the nonlinear nature of Compton scattering represented by the parameter
$\eta \equiv e\sqrt{-\langle A_{\mu} A^{\mu}\rangle}/mc^2$ where $A_{\mu}$ is
the four-vector potential of the laser pulse~\cite{SLACDron}.
The $\eta$ parameter increases the effective electron mass $m^*=m\sqrt{1+\eta^2}$
in the laser field. Hence, the scattered photon energy in the single photon absorption case
is effectively lowered and the photon yield in the Compton edge energy also diminishes.
On the other hand, the larger number of laser photons increases the
luminosity factor in the Compton scattering.
Therefore, there is an optimal $\eta$ value so that the scattered
photon energy is kept within 0.5-1.0 MeV range 
with the maximized $\gamma$-ray yield.

\begin{figure}
\includegraphics[width=0.8\linewidth]{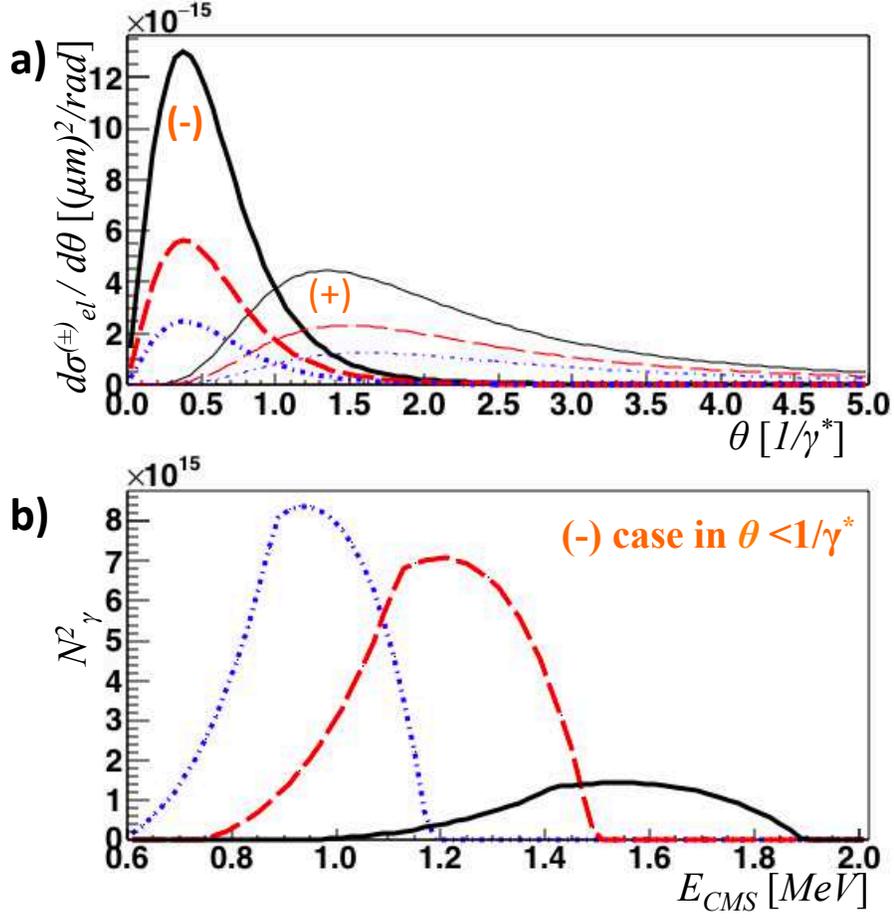}
\caption{ 
Nonlinear effects in Compton scattering
with $\eta = 0.31$(solid), 0.63(dashed) and 0.88(dash-dotted)
:
a) differential Compton scattering cross sections 
$\sigma^{(\pm)}_{el}/d\theta$ as a function of photon scattering
angle measured from the incident direction of the electron beam
in units of $1/\gamma^*$ with the Lorentz factor of the
effective electron mass in the laser field.
The symbols $(\pm)$ denote circular polarization flip $(-)$ and non-flip $(+)$
cases in the transition between the initial and final state photons.
b) $E_{cms}$ distributions by taking all possible
energy combinations between two incident $\gamma$-rays from the Compton scattering
points with the cross section weights in the case of $(-)$ in $\theta < 1/\gamma^*$,
where the distributions are normalized to squares of 
the numbers of generated $\gamma$-rays. 
}
\label{Fig4}
\end{figure}

In Fig.\ref{Fig4}, we plot this situation in three different $\eta$ cases:
solid ($\eta=0.31$), dashed ($\eta=0.63$) and dash-dotted ($\eta=0.88$).
Figure \ref{Fig4} a) shows the differential cross sections
$d\sigma^{(\pm)}_{el}/d\theta$ in units of $(\mu\mbox{m})^2/rad$
as a function of photon scattering
angle measured from the incident direction of the electron beam, 
where the abscissa is displayed in units of $1/\gamma^*$ 
with the Lorentz factor defined by 
the effective electron mass in the laser field.
The symbols $(\pm)$ denote circular polarization flip $(-)$ and non-flip $(+)$
cases in the transition between the initial and final state photons.
The $(-)$ and $(+)$ cases are plotted with thicker and thiner lines, respectively.
The cross sections are numerically calculated based on the expressions
available in Ref.~\cite{SLACDron}. 
In any $\eta$ cases, in order to enhance the purity of the $(-)$ case,  
we must consider $\gamma$-rays produced only in $\theta < 1/\gamma^*$ 
as the effective number of $\gamma$-rays useful for the test of helicity specified
$\gamma\gamma$-scatterings.
Figure \ref{Fig4} b) shows $E_{cms}$ distributions by taking all possible
energy combinations between two incident $\gamma$-rays from the Compton scattering
points with the cross section weights in the $(-)$ case 
within $\theta < 1/\gamma^*$ in Fig.\ref{Fig4} a), 
where the area of the distributions indicate squares of the numbers of generated
$\gamma$-rays, $N_{\gamma}$, corresponding to the numerators in the luminosity factor
for the $\gamma\gamma$-collider as we discuss below.

The head-on luminosity factor for 
$\gamma\gamma$-scattering is defined as
$
%L_{\gamma\gamma}=
\frac{f N^2_{\gamma}}{4\pi r^2_b}
$
where we assume $\gamma\gamma$-collisions take place along the electron 
beam axis within the effective $\gamma$-beam radius at IP, $r_b$. 
Because the distance between the two CP points, $D$, should be kept as small as possible
to increase $N_{\gamma}$,
$r_b$ effectively coincides with the beam waist of the Compton seed laser, 
$w_0$, if the electron beam radius is much smaller than $w_0$. 
On the other hand, in order to enrich the $+ \rightarrow any$ case 
in the QED-based $\gamma\gamma$-scattering, 
the requirement to enhance the $(-)$ case
by limiting the effective scattering angle $\theta < 1/\gamma^*$ demands
the relation $D/\gamma^*=w_0$.
This implies that we can control the purity of the initial circular polarization states
by adjusting $D$ in experiments.

\section{Design of the detection system}
We summarize optimal parameters for an all-optical photon-photon collider 
in the case of $\eta = 0.63$ in Table 1. 
For instance, if we apply this detection system to two 
synchronized lasers operated at $f\sim10$ Hz, a data taking period over 
several months will be sufficient to claim the statistical significance 
of the QED-based elastic scattering events for the $+ \rightarrow any$ case.
 
\renewcommand{\arraystretch}{0.9}
\begin{table}[h!]
\caption{
Parameters for the all-optical photon-photon collider LPA drive laser	
}
\begin{center}
\begin{tabular}{lr}  
\\ \hline
LPA drive laser 
\\ \hline
Wavelength [$\mu$m]	& 0.8\\
Repetition rate [Hz] & 	10\\
Pulse energy [J] & 3.5\\
Peak power [TW] & 41\\
Pulse duration [fs]&85
\\ \hline
LPA electron beam	
\\ \hline
Beam energy [MeV]&210\\
Plasma density [$10^{18}$ cm${}^{-3}$]&1.1\\
Accelerator length [cm]&2.6\\
Charge per bunch [nC]&1.6\\
Bunch duration [fs]& $\sim 10$\\
Normalized emittance [mm mrad]&	$\sim 0.15$\\
Horizontal rms beam size at Compton IP [$\mu$m]&0.8\\
Vertical rms beam size at Compton IP [$\mu$m] &0.5
\\ \hline
Laser for Compton scattering	
\\ \hline
Wavelength [$\mu$m]&0.8\\
Repetition rate [Hz]&10\\
Pulse energy [mJ]&89\\
Pulse duration [fs]&209\\
Spot radius at Compton IP [$\mu$m]&4.0\\
Interaction angle [degree]&0\\
Focused intensity [$10^{17}$ W/cm${}^{2}$]&8.5
\\ \hline
Compton $\gamma$-ray beam	
\\ \hline
Photon energy between $\theta = 0 - 1/\gamma^*$ rad [MeV]&0.37-0.75\\
Effective photon flux [$10^9$ s${}^{-1}$]&8.65\\
Effective spot radius $r_b$ at $\gamma\gamma$-IP [$\mu$m]&4.0
\\ \hline
$\gamma\gamma$-scattering	
\\ \hline
Averaged CMS photon-photon collision energy [MeV]&1.5\\
Averaged QED cross section [$\mu$b]&3.33\\
Rate of QED events [$10^{-6}$ s${}^{-1}$]&2.4\\
\hline
\end{tabular}
\end{center}
\label{Tab1}
\end{table}
\renewcommand{\arraystretch}{1.0}
 
\begin{figure}
\includegraphics[width=1.0\linewidth]{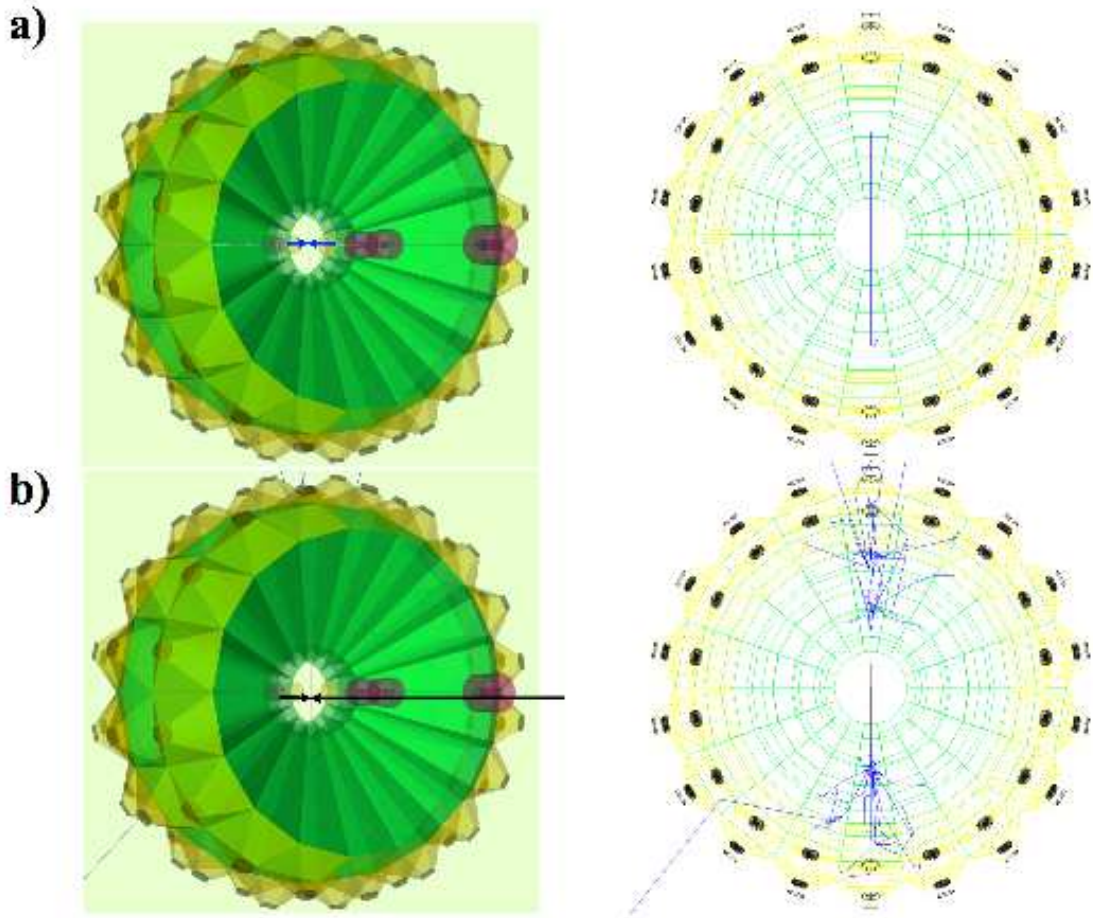}
\caption{ 
Comparison of event displays simulated by GEANT4 for
a) a QED-based $\gamma\gamma\rightarrow\gamma\gamma$ event and
b) a background $e^- e^- \rightarrow e^- e^-$ event, where
the blue and black trajectories denote photons and
electrons, respectively. The detector system covers the polar angle
$\vartheta =45-135$ degree and the azimuthal angle $\varphi=0-360$ degree
around the beam axis consisting of 90 scintillator crystals
(18 crystals in $\varphi \times 5$ layers in $\vartheta$)
made of Ce doped Gd${}_2$SO${}_5$ (GSO:Ce).
GSO:Ce is a well-balanced scintillator from the point of view
on the scintillation-photon yield for a sub-MeV $\gamma$-ray,
the time resolution and the radiation length $X_0$.
The radiation length $X_0=1.38$ cm is reasonably small to
suppress the lateral spreads of the electromagnetic showers produced
by 210 MeV electrons. The individual crystal has the depth of $10 X_0$.
}
\label{Fig7}
\end{figure}

Due to the short $D$, we simultaneously have 
to consider electron-electron scattering after the inverse Compton scattering
occurs,
because this M{\o}ller's scattering can produce dominant background events against 
the elastic $\gamma\gamma$-scattering as shown in Fig.\ref{Fig7} a) and b). 
However, as far as only large angle scattering events in the range of 
$\vartheta=45-135$ degree are measured, the partially integrated cross section 
of M{\o}ller's scattering over that solid angle is suppressed to 17.7 $\mu$b 
for the electron energy of 210 MeV, which is evaluated by the following 
differential cross section with respect to solid angle
%\beqa
$
\frac{d\sigma_M}{d\Omega} = \frac{\alpha^2}{4E^2(E^2-m^2)^2} 
%\times
%\mbox{\hspace{4cm}}
%\\ \nnb
\left[\frac{4(2E^2-m^2)^2}{\sin^4\vartheta}
-\frac{8E^4-4E^2m^2-m^4}{\sin^2\vartheta}+(E^2-m^2)^2 \right]
$
%\eeqa
where $E$ is the incident electron energy in the center-of-mass system. 
On the other hand, the cross section of QED $\gamma\gamma$-scattering is 
relatively enhanced over the background events in the same $\vartheta$ range. 
Taking the higher luminosity factor for the electron-electron scattering 
than that of $\gamma\gamma$-scattering into account, 
the event rate for M{\o}ller's scattering reaches 0.36 Hz for $f=10$ Hz and 
the accidental rate for two types of scattering events to occur within 
the same shot can be evaluated as $2.6\times10^{-6}$ Hz.
This contaminated event rate 
corresponds to 36 \% of that of the QED $\gamma\gamma$-scattering. 
Therefore, even if one throws such contaminated events away without 
the detailed event-by-event offline analysis, 
the statistical loss of the QED events is still acceptable. 
In reality, the energy deposits on the detector 
as well as the event topologies, whether electromagnetic showers exist or not, 
are very different between QED $\gamma\gamma$-scattering and M{\o}ller's 
scattering as displayed in Fig.\ref{Fig7} a) and b). 
Therefore, one can readily distinguish two types of scattering events 
at the offline analysis and possibly distinguish them even within 
the same shot, if the two $\gamma$-clusters are sufficiently 
isolated from the background electromagnetic showers. 
Moreover, counting the number of M{\o}ller's scattering events is 
indispensable to directly measure the electron-electron luminosity 
in the actual experimental condition. This information is essentially important
to deduce the $\gamma\gamma$ luminosity in addition to 
the direct measurement of the $\gamma$-ray flux.

\section{Conclusion}
Verification of the QED-based helicity dependent 
light-by-light scattering 
with the all-optical photon-photon collider is realizable 
in the table-top scale. 
Provided PW-class lasers such as Extreme Light Infrastructure (ELI)~\cite{R31}
we may be able to test the QED-based light-by-light scattering especially in
the same circular polarization case of incident $\gamma$-rays
with an experimentally feasible rate.
Furthermore, the rapid development of fiber-based high-power and high-repetition 
rate lasers at 10~kHz~\cite{R32} would allow us to test the opposite circular polarization
case and enable the comparison between the two helicity configurations 
quantitatively. 
The larger statistics also opens up opportunities to perform general 
resonance searches 
around the MeV energy scale based on the landmark of quantitative 
verification of the QED-based perturbative photon-photon scattering. 
%
%Moreover, by intentionally involving nonlinear Compton scattering processes 
%where much higher photon energies are produced via multi-photon 
%absorption processes, we may be able to extend the search region 
%beyond the MeV range.
%
Finally we emphasize that our proposal corresponds to the first case of LPA-based 
electron-electron and $\gamma\gamma$-colliders applied to 
fundamental particle physics. 
If succeeded, it would be a ground-breaking advancement 
in experimental particle physics 
even if the center of mass energy is still in the MeV range.

%%%%%%%%%%%%
%   ack    %
%%%%%%%%%%%%
\begin{acknowledgments}
We cordially thank Y. Iwashita for deeply considering the possibility to 
introduce the permanent-magnet-based dipole at IP in the electron transport 
line to avoid M{\o}ller's scattering in advance. 
K. Homma appreciates E. Milotti and C. Curceanu for discussions on 
the derivation for the QED-based cross section. 
He also acknowledges for the support by the Collaborative Research
Program of the Institute for Chemical Research,
Kyoto University (grants No. 2015-93).
K. Nakajima was supported by the Project Code (IBS-R012-D1).
\end{acknowledgments}

%%%%%%%%%%%%
%   ref    %
%%%%%%%%%%%%

\end{document}